\documentclass[aps,prl,twocolumn,superscriptaddress,preprintnumbers,longbibliography]{revtex4-1}
\usepackage{amsmath}
\usepackage{amssymb}
\usepackage{graphicx}
\usepackage{gensymb}
\usepackage{float}
\usepackage{physics}
\usepackage{siunitx}
\usepackage{comment}
\usepackage[colorlinks = true, citecolor = blue, linkcolor = blue, urlcolor = blue]{hyperref}
\usepackage{placeins}

\newcommand{\lambdaqcd}{\Lambda_\text{QCD}}

\usepackage[dvipsnames]{xcolor}

\newcommand{\rmi}{{\rm i} }
\newcommand{\rmd}{{\rm d} }

\newcommand{\np}{{\text{np}} }

\begin{document}

\title{Collinear limit of the energy-energy correlator in $e^+ e^-$ collisions: \\
transition from perturbative to non-perturbative regimes}

\author{Enrico Herrmann}
\email{eh10@g.ucla.edu}
\affiliation{Department of Physics and Astronomy, University of California, Los Angeles, CA 90095, USA}
\affiliation{Mani L. Bhaumik Institute for Theoretical Physics, University of California, Los Angeles, CA 90095, USA}

\author{Zhong-Bo Kang}
\email{zkang@physics.ucla.edu}
\affiliation{Department of Physics and Astronomy, University of California, Los Angeles, CA 90095, USA}
\affiliation{Mani L. Bhaumik Institute for Theoretical Physics, University of California, Los Angeles, CA 90095, USA}
\affiliation{Center for Frontiers in Nuclear Science, Stony Brook University, Stony Brook, NY 11794, USA}

\author{Jani Penttala}
\email{janipenttala@physics.ucla.edu}
\affiliation{Department of Physics and Astronomy, University of California, Los Angeles, CA 90095, USA}
\affiliation{Mani L. Bhaumik Institute for Theoretical Physics, University of California, Los Angeles, CA 90095, USA}

\author{Congyue Zhang}
\email{maxzhang2002@g.ucla.edu}
\affiliation{Department of Physics and Astronomy, University of California, Los Angeles, CA 90095, USA}
\affiliation{Mani L. Bhaumik Institute for Theoretical Physics, University of California, Los Angeles, CA 90095, USA}

\begin{abstract}
We study the collinear limit of the energy-energy correlator (EEC) in $e^+e^-$ collisions, focusing on the transition from the perturbative QCD regime at relatively large angles to the non-perturbative region at small angles. To describe this transition, we introduce a non-perturbative jet function and perform a global analysis at NNLO+NNLL accuracy using experimental data spanning center-of-mass energies from $Q = 29.0$ to $91.2$ GeV. This marks the first accurate description of the EEC across the entire near-side region ($0^\circ<\chi<90^\circ$) within a unified theoretical framework.
Our analysis also provides, for the first time, a quantitative extraction of the non-perturbative contribution to the EEC quark jet function, identifying a characteristic transition scale around $2.3$ GeV—distinct from the scale observed in EEC-in-jet measurements in $pp$ collisions at the LHC, which are dominated by gluon jets. These results offer the first evidence for flavor dependence (quark vs.\ gluon) in the EEC and provide new insights into the interplay between perturbative and non-perturbative QCD dynamics.
\end{abstract}

\maketitle 


{\it \textbf{Introduction.}} Energy-flow operators and energy-energy correlators (EEC)~\cite{Basham:1978bw,Basham:1978zq}  have emerged as particularly interesting examples of detector operators~\cite{Kravchuk:2018htv,Caron-Huot:2022eqs,Kologlu:2019mfz} and associated event shapes. They enable detailed studies of Lorentzian properties in general quantum field theories (QFTs) (see e.g.~Ref.~\cite{Hofman:2008ar,Belitsky:2013xxa,Belitsky:2013bja,Belitsky:2013ofa,Kologlu:2019mfz,Chang:2020qpj}), and of QCD dynamics in particular~\cite{Moult:2018jzp,Dixon:2019uzg,Chen:2020adz,Chen:2023zzh}. 
The availability of experimental collider data, along with significant recent theoretical developments---both in formal QFT frameworks and in precision perturbative predictions and resummation---has reignited interest in energy-energy correlators; for recent reviews, see Refs.~\cite{Moult:2025nhu,Neill:2022lqx}. 

The theoretically simplest scenario is the EEC in a state produced by a local operator,
\begin{equation}
\label{eq:EEC_op_def}
\text{EEC} = \frac{ \int \rmd^4x \, e^{\rmi Q\cdot x}\langle 0 | \mathcal{O}^\dagger(x) \mathcal{E}(\hat{n}_1)\mathcal{E}(\hat{n}_2) \mathcal{O}(0) | 0 \rangle}{
\int \rmd^4x \, e^{\rmi Q\cdot x}\langle 0|\mathcal{O}^\dagger(x) \mathcal{O}(0) | 0\rangle }\,,
\end{equation}
where $\hat{n}_i$ are unit vectors indicating the directions of the energy-flow operators $\mathcal{E}$ on the celestial sphere~\cite{Hofman:2008ar}. Experimentally, this setup is realized in $e^+e^-$ annihilation, which is the focus of this work, with the electromagnetic current $J$ playing the role of the local operator~$\mathcal{O}$.   
Complementary to the operator definition in Eq.~(\ref{eq:EEC_op_def}), one can use the action of the energy-flow operators on single-particle states to rewrite the EEC as a weighted cross section; see Eq.~(\ref{eq:EEC_def}) below and Refs.~\cite{Belitsky:2013ofa,Belitsky:2013xxa} for earlier discussions. In collider settings, it is standard to express the EEC in terms of the angle between the pair of detectors, $\hat{n}_1 \cdot \hat{n}_2 = \cos \chi$. The resulting angular dependence of the observable allows the study of interesting kinematical limits. In the \textbf{back-to-back limit} ($\chi\to \pi$), the EEC is sensitive to transverse-momentum dependence of hadronization~\cite{Ebert:2020sfi, Boussarie:2023izj, Kang:2024dja,
Dokshitzer:1978yd, Parisi:1979se,Parisi:1979xd,Kodaira:1981nh,
Chao:1982wb,Soper:1982wc,Kodaira:1982az,Collins:1981va,Collins:1985xx,Collins:1985kw,Collins:1981zc,Kang:2023big}, whereas in the \textbf{collinear} limit ($\chi\to 0$) the EEC can be used to study jet formation \cite{Konishi:1978yx, Konishi:1978ax, Konishi:1979cb}, collinear factorization~\cite{Dixon:2019uzg}, and the light-ray OPE limit \cite{Hofman:2008ar,Kologlu:2019mfz,Chen:2021gdk}. Both of these regions can be computed to high accuracy in perturbation theory using resummation techniques, providing a detailed understanding of the underlying physics.

However, one kinematic regime that remains poorly understood is the extreme collinear region where the relevant momentum scale, $Q\chi$, becomes comparable to the QCD scale parameter $\lambdaqcd$~\cite{Liu:2024lxy,Barata:2024wsu,Fu:2024pic}. In this region, non-perturbative (NP) effects cannot be ignored, and they modify the power-law behavior of the EEC predicted from collinear factorization and jet evolution. This breakdown of the purely perturbative description has been experimentally observed in $e^+ e^-$ collisions~\cite{SLD:1994idb, L3:1992btq, OPAL:1991uui, TOPAZ:1989yod, TASSO:1987mcs, JADE:1984taa, Fernandez:1984db, Wood:1987uf, CELLO:1982rca, PLUTO:1985yzc, OPAL:1990reb, ALEPH:1990vew, L3:1991qlf, SLD:1994yoe, Bossi:2024qeu,Bossi:2025xsi}, and more recently in measurements of the EEC inside jets at hadron colliders~\cite{CMS:2024mlf, ALICE:2024dfl, Komiske:2022enw}, where the transition from the perturbative physics of partons to non-perturbative free hadrons has been clearly observed. This universal behavior of the EEC suggests that the underlying physics in these different experimental environments is similar and should be described by a unified framework. During the completion of our work, Refs.~\cite{Lee:2025okn,Guo:2025zwb,Chang:2025kgq,Kang:2025EEC} suggested a relation between the transition region and dihadron fragmentation, supporting the universality of the non-perturbative physics across different collider systems. 

In this Letter, we focus on the collinear limit of the EEC in $e^+ e^-$ collisions, particularly on the transition from perturbative to non-perturbative QCD regimes. To incorporate non-perturbative contributions, we follow the framework of Ref.~\cite{Barata:2024wsu} and perform the calculation in position space, where non-perturbative physics enters as a multiplicative function---a \emph{NP jet function} multiplying the perturbative expression for the EEC jet function. This approach extends earlier studies~\cite{Korchemsky:1999kt,Lee:2024esz}, which included only the leading power correction and showed improved agreement with data for moderately small angles. However, deep in the non-perturbative regime, a power expansion is insufficient, and the non-perturbative physics must be modeled through a profile function---our non-perturbative jet function. We fit this function to the EEC data, providing a consistent description across different center-of-mass energies in $e^+e^-$ collisions. As the non-perturbative contributions in the collinear limit originate from final-state hadronization, the same non-perturbative function is expected to appear in $pp$ collisions, where only the initial state differs. Our results for $e^+e^-$ collisions can thus be applied in future studies of the EEC in hadron collider environments.


{\it \textbf{Theoretical Framework.}}
In terms of the produced hadrons, the EEC in Eq.~\eqref{eq:EEC_op_def} can also be written as~\cite{Ebert:2020sfi}:
\begin{equation}
\label{eq:EEC_def}
  \dv{\Sigma}{\chi} = \sum_{i, j} \int
  \dd{\sigma_{ij}}
  \frac{E_i E_j}{Q^2} \delta\qty(\chi - \theta_{ij} ).
\end{equation}
This form has the physical interpretation of an energy-weighted probability density to find any pair of final-state hadrons $i$ and $j$ at a specified opening angle $\chi$. Here, $E_i$ and $E_j$ are the hadron energies in the center-of-mass frame of the $e^+e^-$ system, $\theta_{ij}$ is the angle between them, and $Q^2$ is the center-of-mass energy of the collision. The differential $\dd{\sigma_{ij}}$ denotes the semi-inclusive cross section for $e^+e^- \rightarrow i + j + X$.
We also introduce the dimensionless variable $z = \frac12 (1-\cos\chi)$ to streamline the following discussions and write
$\frac{\dd{\Sigma}}{\dd{z}} = \frac{2}{\sin \chi} \frac{\dd{\Sigma}}{\dd{\chi}} $.

Fixed-order perturbative results for the EEC in QCD are available analytically at LO \cite{Basham:1978bw,Basham:1978zq} and NLO \cite{Dixon:2018qgp}, and numerically at NNLO \cite{DelDuca:2016csb,Tulipant:2017ybb}.  In the collinear limit  $z \to 0$ ($\chi \to 0$), the fixed-order expression contains large logarithms of the form $\alpha_s^{\,n}\ln^{m}z$ that must be resummed. To achieve this, we decompose the EEC into a resummed singular term and a finite remainder:
\begin{align}
\frac{\dd{\Sigma}}{\dd{z}} = \frac{\dd{\Sigma^{\rm{res}}}}{\dd{z}} + 
\frac{\dd{\Sigma^{\rm{nons}}}}{\dd{z}} \,,
\end{align}
where $\dd \Sigma^{\rm res}/\dd z$ resums all logarithmically enhanced contributions and $\dd \Sigma^{\rm nons}/\dd z$ collects the non-singular terms that remain finite as $z\!\to\!0$. 

The factorization formalism for resumming $\ln z$ terms in the EEC collinear limit was first derived using jet calculus~\cite{Konishi:1978ax,Konishi:1978yx,Konishi:1979cb,Kalinowski:1980wea,RICHARDS1982193}, valid at leading logarithmic (LL) accuracy. Its extension to next-to-leading logarithmic (NLL) and next-to-next-to-leading logarithmic (NNLL) accuracy was first presented in Ref.~\cite{Dixon:2019uzg}, which we summarize in the supplementary material. In the extreme collinear limit, where $\sqrt{z} Q$ is of order $\Lambda_{\text{QCD}}$, the EEC becomes sensitive to small transverse momentum generated during hadronization of the outgoing partons, such as non-perturbative soft radiation, in addition to the collinear emission described by perturbative factorization and jet evolution. These non-perturbative transverse-momentum effects can be incorporated via a convolution with the perturbative result in transverse momentum. To simplify the implementation, it is much easier to handle this in position space, where the convolution becomes a simple product~\cite{Collins:2011zzd}. We introduce the transverse coordinate $b$, which is Fourier-conjugate to the transverse momentum scale $q_T=\sqrt{z}\, Q=\sin(\chi/2)\, Q$. A detailed derivation of this position-space representation is provided in the supplementary material. Concretely, the resummed term of the EEC can be written as
\begin{align}
\label{eq:factorization}
\frac{1}{\hat{\sigma}_0} \frac{\dd{\Sigma^{\rm res}}}{\dd{z}} = \,
& \frac{Q^{2}}{2} \int_{0}^{\infty} \dd{b} \, b\, J_{0}(b\, q_T) \, 
 \\
&\times \int_{0}^{1} \dd{x}\,x^{2}\,
  \tilde{\mathbf j}\qty(\frac{b}{x},\mu) \vdot
  \mathbf{H}(x,Q,\mu)\, , \nonumber
\end{align}
where $x$ is the energy fraction of the outgoing parton initiating the jet. This expression is factorized into a convolution of the two-component EEC differential jet function $\tilde{\mathbf{j}}=(\tilde{j}_q,\,\tilde{j}_g)$, representing quark and gluon jets, and the hard function $\mathbf{H}=(H_q,\,H_g)$, normalized to the Born cross section $\hat{\sigma}_0$ for $e^+e^- \rightarrow q\bar{q}$. 

The hard and jet functions obey renormalization group (RG) equations~\cite{Dixon:2019uzg}
\begin{equation}
\begin{aligned}
\hspace{-.5cm}
\frac{\dd}{\dd\ln\mu^{2}}\,\mathbf{H}(x,Q,\mu)
  &  = -\int_x^1 \frac{\dd{y}}{y}\,
      \mathbf{P}(y,\mu)\vdot
      \mathbf{H}\!\left(\frac{x}{y},Q,\mu\right),
\hspace{-.5cm}\\
\hspace{-.5cm}
\frac{\dd}{\dd\ln\mu^{2}}\,
\widetilde{\mathbf{j}}(b,\mu)
   & = \int_{0}^{1} \dd{y}\,y^{2}\,
      \widetilde{\mathbf{j}}\!\left(\frac{b}{y},\mu\right)\vdot
      \mathbf{P}(y,\mu),
\hspace{-.5cm}      
\end{aligned}
\label{eq:J-H-RG}
\end{equation}
where the jet function in position space 
obeys the same RG evolution as its momentum-space counterpart (see details in the supplementary material). The time-like splitting function
$
\mathbf{P}
= \begin{pmatrix}
P_{qq} & P_{qg}\\
P_{gq} & P_{gg}
\end{pmatrix}
$
was computed to NNLO~\cite{Rijken:1996ns, Mitov:2006wy, Mitov:2006ic, Moch:2007tx, Almasy:2011eq}. The canonical choice of scales that eliminates large logarithms in the initial conditions for the hard and jet functions is $\mu_H^i  = Q$ and
$\mu_J^i  = \mu_{b_\ast}$, respectively, where $\mu_{b_\ast}=2e^{-\gamma_E}/b_\ast$. We regulate the Landau pole using the standard $b_*$ prescription~\cite{Collins:1984kg, Collins:2014jpa}, where $b_*=b/\sqrt{1+b^2/b_{\rm max}^2}$, and choose $b_{\rm max} = 2e^{-\gamma_E} \,\mathrm{GeV}^{-1}$. In our central prediction, we evolve the jet functions from their canonical scale $\mu_J^i$ up to the common renormalization scale $\mu=Q$ to resum the large logarithms in the collinear limit, while no evolution is needed for the hard function.

In the small-$b$ region, where $1/b \gg \Lambda_{\text{QCD}}$, the jet functions can be computed perturbatively, and we derive their expressions in position space at NNLO using their known forms in momentum space in the supplementary material. 
However, in the large-$b$ region, the jet function receives non-perturbative contributions from hadronization.
To incorporate these effects, we multiply the perturbative jet function by a non-perturbative correction at the initial jet scale $\mu_J^i  = \mu_{b_\ast}$,
\begin{align}
   \tilde{j}_{i}(b,\mu_{b_\ast}) =  \tilde{j}_{i,\mathrm{pert}}(b_*,\mu_{b_\ast}) \, j_{i,\np}(b)\,,
\end{align}
with $i = q, g$. Then, the resummed jet function $\tilde{j}_{i}(b,\mu)$ is evolved to the scale $\mu$ by solving the jet RG equation in Eq.~\eqref{eq:J-H-RG} order by order up to $\mathcal{O}(\alpha_s^9)$.

In principle, one would expect different non-perturbative contributions for quark and gluon jets. For example, \textsc{Pythia} simulations~\cite{Liu:2024lxy,Apolinario:2025vtx} have shown that the EEC distribution of a gluon jet is broader, with the peak marking the transition from the perturbative to the non-perturbative regime occurring at a larger angle than for a quark jet. However, since the EEC in $e^+e^-$ collisions is dominated by quark jets, we find that we cannot separately constrain the non-perturbative jet functions for quark and gluon jets. We therefore adopt the same function for both, $j_{q,\np}(b) = j_{g,\np}(b) \equiv j_{\np}(b)$. In addition, to ensure the correct perturbative limit, we impose the condition $\lim_{b \to 0} j_{i,\np}(b) = 1$. Inspired by our earlier study of the EEC inside jets in $pp$ collisions~\cite{Barata:2024wsu} and in the back-to-back limit~\cite{Kang:2024dja}, we use the ansatz
\begin{equation}
j_\np(b) = \exp\left[ - (a_1\,b)^{a_2} \right],
\end{equation}
where $a_1$ and $a_2$ are free parameters fitted to the experimental data. Notably, when $a_2 \simeq 1 \ (\delta a_2 = a_2 - 1 \ll 1)$, the resummed EEC at $\order{\alpha_s^0}$ equals
\begin{equation}
\frac{1}{\hat{\sigma}_0} \frac{\dd{\Sigma^{\rm{res}}}}{\dd{z}} = \frac{1}{2 \, z^{3/2}} \frac{{a_1}/{4}}{Q} + \order{\frac{1}{Q^3}} + \order{\delta a_2} + \order{\alpha_s},
\end{equation}
which reproduces the leading power correction to the EEC in the relatively small-$z$ limit~\cite{Korchemsky:1999kt, Lee:2024esz}.

The non-singular contribution can be expressed as
\begin{align}
\frac{\dd{\Sigma^{\rm{nons}}}}{\dd{z}} = 
    \frac{\dd{\Sigma^{\rm{F.O.}}}}{\dd{z}} 
-   \frac{\dd{\Sigma^{\rm{sing}}}}{\dd{z}} \, .
\end{align}
Here, $\dd \Sigma^{\rm{F.O.}}/\dd z$ refers to the full fixed-order EEC, for which we use the available analytic results at LO~\cite{Basham:1978bw} and NLO~\cite{Dixon:2018qgp}, together with the numerical result at NNLO~\cite{DelDuca:2016csb}. The singular contribution $\dd \Sigma^{\rm{sing}}/\dd z$ before resummation is obtained by truncating the resummed EEC to $\order{\alpha_{s}^{3}}$, which is then matched to the NNLO fixed-order EEC. 

To account for missing higher-order corrections, we include an additional $K$-factor $B$ that multiplies this non-singular contribution in our fits. We also include a $K$-factor $A$ for the resummed term $\Sigma^{\rm{res}}$. In total, we have four free parameters, $a_1$, $a_2$, $A$, and $B$, which are determined by fitting to data.

\begin{table}[t!]
\centering
\begin{tabular}{|c c c c|}
\hline
\hline
Collaboration & $Q$\,(\rm{GeV}) & $N_\mathrm{data}$ & $\chi^2$  \\
\hline
OPAL~\cite{OPAL:1991uui}    & 91.2 & 89 & 100 \\
SLD~\cite{SLD:1994idb}      & 91.2 & 24 & 30.5\\
TOPAZ~\cite{TOPAZ:1989yod}  & 59.5 & 24 & 36.8\\
TOPAZ~\cite{TOPAZ:1989yod}  & 53.3 & 24 & 60.9\\
TASSO~\cite{TASSO:1987mcs}  & 43.5 & 24 & 37.0\\
TASSO~\cite{TASSO:1987mcs}  & 34.8 & 24 & 16.9\\
MARK II~\cite{Wood:1987uf}   & 29.0 & 24 & 35.4\\
MAC~\cite{Fernandez:1984db} & 29.0 & 24 & 35.9\\
Total                       &      & 257 & 354\\
\hline
\hline
\end{tabular}
\caption{Experimental datasets used in the fits and the corresponding $\chi^2$ values for the best-fit results.}
\label{table1}
\end{table}

\begin{table}[htb]
\centering
\renewcommand{\arraystretch}{1.6}
\setlength{\tabcolsep}{6pt}  

\begin{tabular}{|ll|}
\hline
$A = 0.858^{+0.002}_{-0.003}{}^{+0.001}_{-0.002}$ &
$B = 1.238^{+0.010}_{-0.010}{}^{+0.009}_{-0.016}$ \\
$a_1 = 2.310^{+0.015}_{-0.010}{}^{+0.248}_{-0.279}\,\mathrm{GeV}$ &
$a_2 = 1.059^{+0.006}_{-0.006}{}^{+0.010}_{-0.036}$ \\
\hline
\end{tabular}

\caption{Fitted parameter values with statistical and theoretical uncertainties (68\% confidence level).}
\label{tab:Params}
\end{table}

{\it \textbf{Global Analysis.}} {\it Data Selection.} 
Firstly, to avoid sizable finite-quark-mass corrections that appear at low energies, we retain only datasets with $Q\ge 29.0~\text{GeV}$, a threshold motivated by another EEC study~\cite{Kardos:2018kqj}. Secondly, our factorization formalism is valid only when all final-state hadrons---charged and neutral---are included in the measurement, and therefore we exclude datasets limited to charged tracks. 
Finally, we require complete uncertainty information (statistical and systematic, or their quadratic combination) and sufficient angular resolution (at least 50 bins over the full \(\chi\) range).  
Applying these criteria leaves eight datasets from OPAL \cite{OPAL:1991uui}, SLD \cite{SLD:1994idb}, TOPAZ \cite{TOPAZ:1989yod}, TASSO \cite{TASSO:1987mcs}, MARK II \cite{Wood:1987uf}, and MAC~\cite{Fernandez:1984db}, spanning \(29.0~\text{GeV} \le Q \le 91.2~\text{GeV}\). We summarize these datasets in Table~\ref{table1}. We restrict the fit range to near-side $(0^{\circ} < \chi < 90^{\circ})$, while excluding the first point of each dataset where the measurement is inaccurate. Finally, for measurements whose uncertainties were rounded to 1 significant digit, we expand them to 2 significant digits assuming maximal uncertainty before rounding, following the procedure in Ref.~\cite{Kardos:2018kqj}.

\begin{figure*}[t!]
    \centering
    \includegraphics[width =0.9\textwidth]{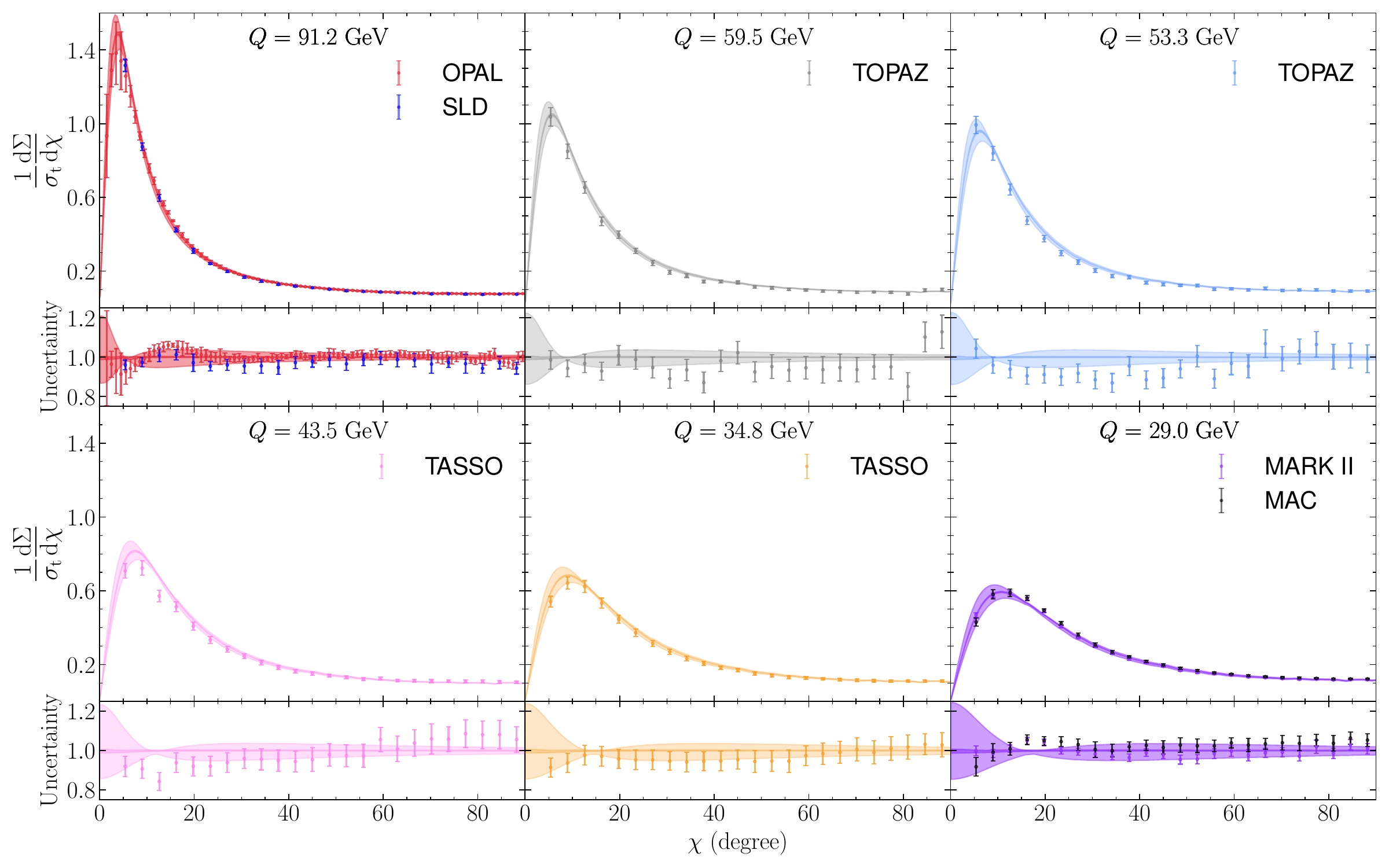} 
    \caption{Comparison of the EEC theory prediction with experimental data. 
    The light and  dark bands represent the 68\% theoretical uncertainty and the 68\% fit uncertainty, respectively.
    The total cross section $\sigma_{\rm{t}}$ is calculated at 2-loop~\cite{Kardos:2018kqj}.
    }
    \label{FIG1}
\end{figure*}

{\it Fit Method.} 
The four free parameters, $A$, $B$, $a_{1}$, and $a_{2}$, are determined by a global $\chi^{2}$ fit.  
For any trial parameter vector $\mathbf p$ we define
$\chi^2\left(\mathbf{p}\right) = \sum_{i} \left(T_i\left(\mathbf{p}\right)-E_i\right)^2 / {\sigma_i^2}$, 
where the sum runs over every fitted experimental data point.  
Here \(E_{i}\) is the measured EEC value,  
\(T_{i}(\mathbf p)\) is the corresponding theoretical prediction obtained with parameters \(\mathbf p\),  
and \(\sigma_{i}\) is the total uncertainty obtained by combining statistical, systematic, and numerical EEC NNLO errors in quadrature. The optimal parameter set is found by minimizing \(\chi^{2}(\mathbf p)\).


{\it Treatment of Uncertainties.}
To estimate the fit uncertainty, we propagate the uncertainty in experimental data points into the fit parameters
$\{A,B,a_1,a_2\}$  
via the Monte-Carlo replica technique \cite{replica-spin-class,Ball:2008by}.  
An ensemble of \(N_{\text{rep}}=200\) replica-datasets is generated by shifting every data point by a Gaussian random noise with width equal to the corresponding experimental error. Each replica is then refitted, producing 200 best-fit parameter vectors. For each parameter, we define the ensemble median as its central value, while the 16th and 84th percentiles' deviation from the median gives lower and upper fit error.

The theory uncertainty due to unknown higher-order corrections to jet functions is assessed via scale variation as in Ref.~\cite{Bell:2023dqs}. Starting from the canonical choice for $\mu_J^{i}$, we multiply it by an independent ratio drawn uniformly from the interval $[1/2,2]$ with 0.1 spacing and refit the data. The 21 scale choices yield a second ensemble of parameter sets. Similarly, the 16th to 84th percentile range of that ensemble is quoted as the theory uncertainty.


{\it Numerical Results.}
The fitted central parameters and their $68\,\%$ uncertainties are collected in Table~\ref{tab:Params}, and the corresponding predictions for the EEC are shown in Fig.~\ref{FIG1} in comparison with the experimental data. The goodness of the fit is given by $\chi^{2}/N_{\mathrm{data}} = 1.38$, which indicates that our fit agrees well with the data. 

The fitted $K$-factors, $A {\sim} 0.86$, and $B {\sim} 1.24$ in Table~\ref{tab:Params}, indicate a sensitivity of the EEC to higher-order corrections including a non-trivial $z$-dependence that cannot be fully captured by constant $K$-factors. Assessing the impact of ${\rm N^3LO}$ fixed-order and higher-order resummation results on our fits, once available, would be an important direction for future work.

\begin{figure*}[t]
    \centering
    \hspace*{-1cm}\includegraphics[width =1.1\textwidth]{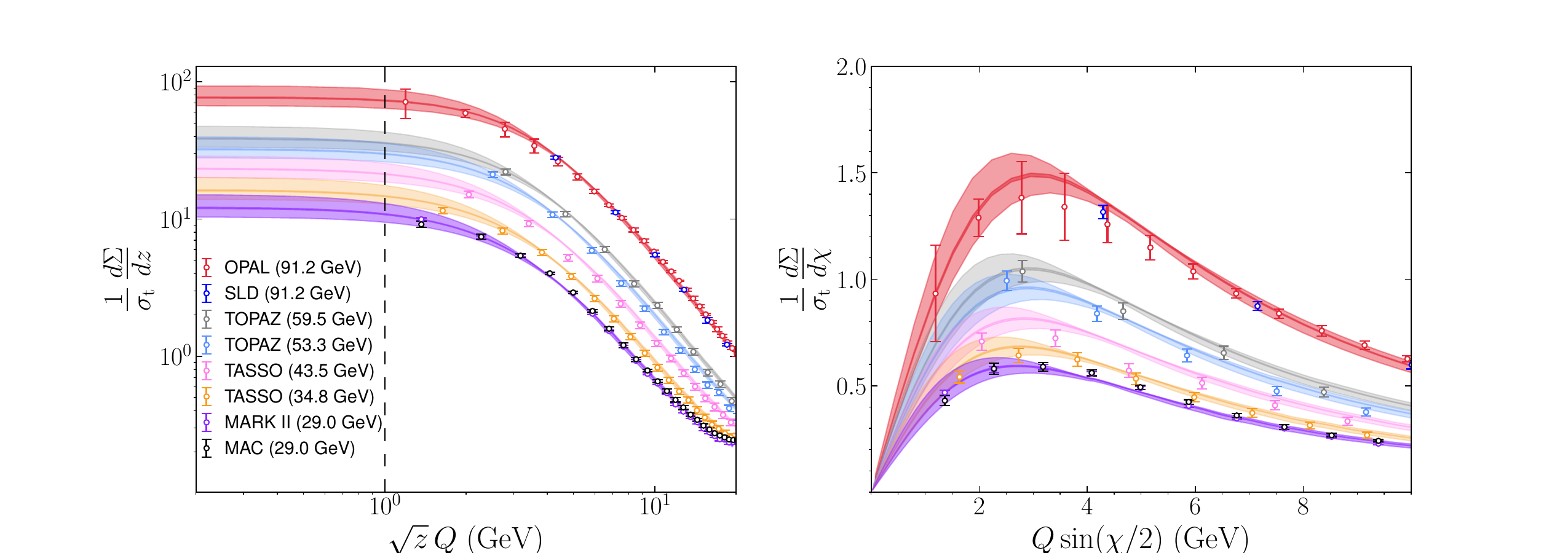} 
    \caption{
    Combined comparison of the EEC theory prediction with experimental data. The light and  dark bands represent the 68\% theoretical uncertainty and the 68\% fit uncertainty, respectively. We observe a characteristic transition scale from perturbative to non-perturbative jet dynamics of $2.3$ GeV.
    }
    \label{FIG2}
\end{figure*}

We also observe that the fit favors a nearly-linear $b$-dependence, with $a_2 = 1.06$. This agrees with our previous finding that a linear $b$-dependence can successfully describe the EEC in the transition region for jets in $pp$ collisions at the LHC~\cite{Barata:2024wsu}.
The left panel in Fig.~\ref{FIG2} shows our prediction for the EEC as a function of the transverse momentum $q_T =\sqrt{z}\,Q$. The vertical dashed line marks $q_T \simeq 1~\text{GeV}$, below which the transverse momenta of the hadrons are of the order $\Lambda_{\rm QCD}$; here, the EEC enters the free-hadron region where confinement erases partonic correlations. Although the free-hadron region ($q_T\lesssim 1$ GeV) is outside our fit range due to lack of experimental data, our prediction for the EEC flattens, exhibiting the expected geometric scaling $\mathrm{d}\Sigma/\mathrm{d}z \sim z^{0}$, see e.g.~\cite{Moult:2025nhu}.
It is also interesting to note that this exponent is close to the value $a_{2}=1.15$ extracted from the back-to-back EEC analysis of Ref.~\cite{Kang:2024dja}, suggesting a universal NP modification across the two limits of the EEC, as both characterize the transition from perturbative partons to non-perturbative hadrons. Further studies along this direction would enrich our understanding of hadronization and clarify the connection between these two regions.

Finally, the coefficient $a_{1}=2.31~\text{GeV}$ is a characteristic scale for the transition between the free hadron region and the partonic region. To illustrate this, the right panel of Fig.~\ref{FIG2} shows the EEC $\mathrm{d}\Sigma/\mathrm{d}\chi$ as a function of the transverse momentum $Q\sin(\chi/2)$. One observes that the transition peak occurs around $2.8~\text{GeV}$, comparable to $a_{1}$.
Because the EEC in $e^{+}e^{-}$ collisions is quark jet dominated, this implies a quark jet transition scale around $2.3$~GeV. This can be compared to the EEC measurements for jets in $pp$ collisions at the LHC, where, because of the much higher energies, the EEC is mostly dominated by gluon jets. The characteristic scale for inclusive jets obtained from a LL fit to the CMS data~\cite{CMS:2024mlf} is $a_{1}=3.8~\mathrm{GeV}$~\cite{Barata:2024wsu}, indicating a larger transition scale for gluon jets, consistent with recent \textsc{Pythia} studies~\cite{Liu:2024lxy,Apolinario:2025vtx}.
A future combined NNLL fit of the $e^{+} e^{-}$ and LHC $pp$ datasets would be very useful, offering the opportunity to separately constrain the NP contributions of quark and gluon jets, thereby allowing us to better understand the flavor (quark vs.\ gluon) dependence of hadronization.

{\it \textbf{Conclusions.}}
In summary, we have carried out a global analysis of the EEC in $e^+e^-$ annihilation at NNLO+NNLL accuracy, covering the full near-side angular region ($0^\circ < \chi < 90^\circ$) and a wide range of center-of-mass energies ($Q = 29.0$--$91.2$ GeV). To describe the transition from perturbative to non-perturbative dynamics, we introduced a non-perturbative modification to the quark jet function and extracted its parameters through a global fit to the data. This analysis identifies a characteristic transition scale around $2.3$ GeV, significantly lower than the scale ($3.8$ GeV) extracted from gluon-dominated EEC-in-jet measurements in $pp$ collisions~\cite{Barata:2024wsu}, thereby providing the first direct evidence of flavor dependence in the EEC. The functional form of the extracted non-perturbative contribution is similar in shape to that found in $pp$ collisions, suggesting a degree of universality, while the shift in scale reflects differences in the underlying partonic configurations and their sensitivity to hadronization effects in quark- versus gluon-initiated processes.

Our approach can be naturally extended to track-based EEC observables~\cite{Jaarsma:2023ell,Lee:2023npz,Lee:2023tkr}, which offer enhanced angular resolution in the small-angle region by leveraging the capabilities of modern tracking systems. For instance, the recent reanalysis of ALEPH data in $e^+e^-$ collisions~\cite{Bossi:2025xsi} provides an opportunity to study the non-perturbative transition using this framework. Similar measurements have been performed in $pp$ collisions by the STAR collaboration at RHIC~\cite{STAR:2025jut} and the ALICE collaboration at the LHC~\cite{ALICE:2024dfl}. Extending EEC studies to $ep$ collisions—for example, using archival data from HERA or future measurements at the Electron-Ion Collider—would further test the universality of hadronization effects and deepen our understanding of non-perturbative QCD.

{\it Acknowledgments.}
We thank G\'abor Somogyi for providing the numerical NNLO EEC results from~\cite{Tulipant:2017ybb,DelDuca:2016csb}. This work is supported by the U.S. Department of Energy (DOE) under award number DE-SC0009937 (E.H.) and by the National Science Foundation under grant No.~PHY-1945471 (Z.K., J.P., and C.Z.). It is also supported by the U.S. Department of Energy, Office of Science, Office of Nuclear Physics, within the framework of the Saturated Glue (SURGE) Topical Theory Collaboration. We are also grateful to the Mani L. Bhaumik Institute for Theoretical Physics for support.
 
\FloatBarrier 

\bibliographystyle{JHEP-2modlong.bst}
\bibliography{references}

\providecommand{\href}[2]{#2}\begingroup\raggedright\begin{thebibliography}{10}

\bibitem{Basham:1978bw}
C.~L. Basham, L.~S. Brown, S.~D. Ellis and S.~T. Love, {\it {Energy Correlations in electron - Positron Annihilation: Testing QCD}},  \href{http://dx.doi.org/10.1103/PhysRevLett.41.1585}{{\em Phys. Rev. Lett.} {\bf 41} (1978) 1585}.

\bibitem{Basham:1978zq}
C.~L. Basham, L.~S. Brown, S.~D. Ellis and S.~T. Love, {\it {Energy Correlations in electron-Positron Annihilation in Quantum Chromodynamics: Asymptotically Free Perturbation Theory}},  \href{http://dx.doi.org/10.1103/PhysRevD.19.2018}{{\em Phys. Rev. D} {\bf 19} (1979) 2018}.

\bibitem{Kravchuk:2018htv}
P.~Kravchuk and D.~Simmons-Duffin, {\it {Light-ray operators in conformal field theory}},  \href{http://dx.doi.org/10.1007/JHEP11(2018)102}{{\em JHEP} {\bf 11} (2018) 102} [\href{http://arXiv.org/abs/1805.00098}{{\tt arXiv:1805.00098 [hep-th]}}].

\bibitem{Caron-Huot:2022eqs}
S.~Caron-Huot, M.~Kologlu, P.~Kravchuk, D.~Meltzer and D.~Simmons-Duffin, {\it {Detectors in weakly-coupled field theories}},  \href{http://dx.doi.org/10.1007/JHEP04(2023)014}{{\em JHEP} {\bf 04} (2023) 014} [\href{http://arXiv.org/abs/2209.00008}{{\tt arXiv:2209.00008 [hep-th]}}].

\bibitem{Kologlu:2019mfz}
M.~Kologlu, P.~Kravchuk, D.~Simmons-Duffin and A.~Zhiboedov, {\it {The light-ray OPE and conformal colliders}},  \href{http://dx.doi.org/10.1007/JHEP01(2021)128}{{\em JHEP} {\bf 01} (2021) 128} [\href{http://arXiv.org/abs/1905.01311}{{\tt arXiv:1905.01311 [hep-th]}}].

\bibitem{Hofman:2008ar}
D.~M. Hofman and J.~Maldacena, {\it {Conformal collider physics: Energy and charge correlations}},  \href{http://dx.doi.org/10.1088/1126-6708/2008/05/012}{{\em JHEP} {\bf 05} (2008) 012} [\href{http://arXiv.org/abs/0803.1467}{{\tt arXiv:0803.1467 [hep-th]}}].

\bibitem{Belitsky:2013xxa}
A.~V. Belitsky, S.~Hohenegger, G.~P. Korchemsky, E.~Sokatchev and A.~Zhiboedov, {\it {From correlation functions to event shapes}},  \href{http://dx.doi.org/10.1016/j.nuclphysb.2014.04.020}{{\em Nucl. Phys. B} {\bf 884} (2014) 305} [\href{http://arXiv.org/abs/1309.0769}{{\tt arXiv:1309.0769 [hep-th]}}].

\bibitem{Belitsky:2013bja}
A.~V. Belitsky, S.~Hohenegger, G.~P. Korchemsky, E.~Sokatchev and A.~Zhiboedov, {\it {Event shapes in $\mathcal{N} = 4$ super-Yang-Mills theory}},  \href{http://dx.doi.org/10.1016/j.nuclphysb.2014.04.019}{{\em Nucl. Phys. B} {\bf 884} (2014) 206} [\href{http://arXiv.org/abs/1309.1424}{{\tt arXiv:1309.1424 [hep-th]}}].

\bibitem{Belitsky:2013ofa}
A.~V. Belitsky, S.~Hohenegger, G.~P. Korchemsky, E.~Sokatchev and A.~Zhiboedov, {\it {Energy-Energy Correlations in N=4 Supersymmetric Yang-Mills Theory}},  \href{http://dx.doi.org/10.1103/PhysRevLett.112.071601}{{\em Phys. Rev. Lett.} {\bf 112} (2014)~no.~7 071601} [\href{http://arXiv.org/abs/1311.6800}{{\tt arXiv:1311.6800 [hep-th]}}].

\bibitem{Chang:2020qpj}
C.-H. Chang, M.~Kologlu, P.~Kravchuk, D.~Simmons-Duffin and A.~Zhiboedov, {\it {Transverse spin in the light-ray OPE}},  \href{http://dx.doi.org/10.1007/JHEP05(2022)059}{{\em JHEP} {\bf 05} (2022) 059} [\href{http://arXiv.org/abs/2010.04726}{{\tt arXiv:2010.04726 [hep-th]}}].

\bibitem{Moult:2018jzp}
I.~Moult and H.~X. Zhu, {\it {Simplicity from Recoil: The Three-Loop Soft Function and Factorization for the Energy-Energy Correlation}},  \href{http://dx.doi.org/10.1007/JHEP08(2018)160}{{\em JHEP} {\bf 08} (2018) 160} [\href{http://arXiv.org/abs/1801.02627}{{\tt arXiv:1801.02627 [hep-ph]}}].

\bibitem{Dixon:2019uzg}
L.~J. Dixon, I.~Moult and H.~X. Zhu, {\it {Collinear limit of the energy-energy correlator}},  \href{http://dx.doi.org/10.1103/PhysRevD.100.014009}{{\em Phys. Rev. D} {\bf 100} (2019)~no.~1 014009} [\href{http://arXiv.org/abs/1905.01310}{{\tt arXiv:1905.01310 [hep-ph]}}].

\bibitem{Chen:2020adz}
H.~Chen, I.~Moult and H.~X. Zhu, {\it {Quantum Interference in Jet Substructure from Spinning Gluons}},  \href{http://dx.doi.org/10.1103/PhysRevLett.126.112003}{{\em Phys. Rev. Lett.} {\bf 126} (2021)~no.~11 112003} [\href{http://arXiv.org/abs/2011.02492}{{\tt arXiv:2011.02492 [hep-ph]}}].

\bibitem{Chen:2023zzh}
H.~Chen, {\it {QCD factorization from light-ray OPE}},  \href{http://dx.doi.org/10.1007/JHEP01(2024)035}{{\em JHEP} {\bf 01} (2024) 035} [\href{http://arXiv.org/abs/2311.00350}{{\tt arXiv:2311.00350 [hep-ph]}}].

\bibitem{Moult:2025nhu}
I.~Moult and H.~X. Zhu, {\it {Energy Correlators: A Journey From Theory to Experiment}},  \href{http://arXiv.org/abs/2506.09119}{{\tt arXiv:2506.09119 [hep-ph]}}.

\bibitem{Neill:2022lqx}
D.~Neill, G.~Vita, I.~Vitev and H.~X. Zhu in {\em {Snowmass 2021}}, 3, 2022.
\newblock \href{http://arXiv.org/abs/2203.07113}{{\tt arXiv:2203.07113 [hep-ph]}}.

\bibitem{Ebert:2020sfi}
M.~A. Ebert, B.~Mistlberger and G.~Vita, {\it {The Energy-Energy Correlation in the back-to-back limit at N$^{3}$LO and N$^{3}$LL'}},  \href{http://dx.doi.org/10.1007/JHEP08(2021)022}{{\em JHEP} {\bf 08} (2021) 022} [\href{http://arXiv.org/abs/2012.07859}{{\tt arXiv:2012.07859 [hep-ph]}}].

\bibitem{Boussarie:2023izj}
R.~Boussarie {\em et.~al.}, {\it {TMD Handbook}},  \href{http://arXiv.org/abs/2304.03302}{{\tt arXiv:2304.03302 [hep-ph]}}.

\bibitem{Kang:2024dja}
Z.-B. Kang, J.~Penttala and C.~Zhang, {\it {Determination of the strong coupling constant and the Collins-Soper kernel from the energy-energy correlator in $e^+e^-$ collisions}},  \href{http://arXiv.org/abs/2410.21435}{{\tt arXiv:2410.21435 [hep-ph]}}.

\bibitem{Dokshitzer:1978yd}
Y.~L. Dokshitzer, D.~Diakonov and S.~I. Troian, {\it {On the Transverse Momentum Distribution of Massive Lepton Pairs}},  \href{http://dx.doi.org/10.1016/0370-2693(78)90240-X}{{\em Phys. Lett. B} {\bf 79} (1978) 269}.

\bibitem{Parisi:1979se}
G.~Parisi and R.~Petronzio, {\it {Small Transverse Momentum Distributions in Hard Processes}},  \href{http://dx.doi.org/10.1016/0550-3213(79)90040-3}{{\em Nucl. Phys. B} {\bf 154} (1979) 427}.

\bibitem{Parisi:1979xd}
G.~Parisi, {\it {Summing Large Perturbative Corrections in QCD}},  \href{http://dx.doi.org/10.1016/0370-2693(80)90746-7}{{\em Phys. Lett. B} {\bf 90} (1980) 295}.

\bibitem{Kodaira:1981nh}
J.~Kodaira and L.~Trentadue, {\it {Summing Soft Emission in QCD}},  \href{http://dx.doi.org/10.1016/0370-2693(82)90907-8}{{\em Phys. Lett. B} {\bf 112} (1982) 66}.

\bibitem{Chao:1982wb}
S.-C. Chao, D.~E. Soper and J.~C. Collins, {\it {The Order $\alpha^- s^2$ Energy-energy Correlation Function at Small Angles}},  \href{http://dx.doi.org/10.1016/0550-3213(83)90248-1}{{\em Nucl. Phys. B} {\bf 214} (1983) 513}.

\bibitem{Soper:1982wc}
D.~E. Soper, {\it {ANGULAR DISTRIBUTION OF TWO OBSERVED HADRONS IN ELECTRON - POSITRON ANNIHILATION}},  \href{http://dx.doi.org/10.1007/BF01571904}{{\em Z. Phys. C} {\bf 17} (1983) 367}.

\bibitem{Kodaira:1982az}
J.~Kodaira and L.~Trentadue, {\it {Single Logarithm Effects in electron-Positron Annihilation}},  \href{http://dx.doi.org/10.1016/0370-2693(83)91213-3}{{\em Phys. Lett. B} {\bf 123} (1983) 335}.

\bibitem{Collins:1981va}
J.~C. Collins and D.~E. Soper, {\it {Back-To-Back Jets: Fourier Transform from B to K-Transverse}},  \href{http://dx.doi.org/10.1016/0550-3213(82)90453-9}{{\em Nucl. Phys. B} {\bf 197} (1982) 446}.

\bibitem{Collins:1985xx}
J.~C. Collins and D.~E. Soper, {\it {The Two Particle Inclusive Cross-section in $e^+ e^-$ Annihilation at {PETRA}, {PEP} and {LEP} Energies}},  \href{http://dx.doi.org/10.1016/0550-3213(87)90035-6}{{\em Nucl. Phys. B} {\bf 284} (1987) 253}.

\bibitem{Collins:1985kw}
J.~C. Collins and D.~E. Soper, {\it {Transverse Momentum in $e^+ e^- \to$ a + $B$ + X}},  {\em Acta Phys. Polon. B} {\bf 16} (1985) 1047.

\bibitem{Collins:1981zc}
J.~C. Collins and D.~E. Soper, {\it {Back-to-back Jets in {QCD}: Comparison With Experiment}},  \href{http://dx.doi.org/10.1103/PhysRevLett.48.655}{{\em Phys. Rev. Lett.} {\bf 48} (1982) 655}.

\bibitem{Kang:2023big}
Z.-B. Kang, K.~Lee, D.~Y. Shao and F.~Zhao, {\it {Probing transverse momentum dependent structures with azimuthal dependence of energy correlators}},  \href{http://dx.doi.org/10.1007/JHEP03(2024)153}{{\em JHEP} {\bf 03} (2024) 153} [\href{http://arXiv.org/abs/2310.15159}{{\tt arXiv:2310.15159 [hep-ph]}}].

\bibitem{Konishi:1978yx}
K.~Konishi, A.~Ukawa and G.~Veneziano, {\it {A Simple Algorithm for QCD Jets}},  \href{http://dx.doi.org/10.1016/0370-2693(78)90015-1}{{\em Phys. Lett. B} {\bf 78} (1978) 243}.

\bibitem{Konishi:1978ax}
K.~Konishi, A.~Ukawa and G.~Veneziano, {\it {On the Transverse Spread of QCD Jets}},  \href{http://dx.doi.org/10.1016/0370-2693(79)90212-0}{{\em Phys. Lett. B} {\bf 80} (1979) 259}.

\bibitem{Konishi:1979cb}
K.~Konishi, A.~Ukawa and G.~Veneziano, {\it {Jet Calculus: A Simple Algorithm for Resolving QCD Jets}},  \href{http://dx.doi.org/10.1016/0550-3213(79)90053-1}{{\em Nucl. Phys. B} {\bf 157} (1979) 45}.

\bibitem{Chen:2021gdk}
H.~Chen, I.~Moult and H.~X. Zhu, {\it {Spinning gluons from the QCD light-ray OPE}},  \href{http://dx.doi.org/10.1007/JHEP08(2022)233}{{\em JHEP} {\bf 08} (2022) 233} [\href{http://arXiv.org/abs/2104.00009}{{\tt arXiv:2104.00009 [hep-ph]}}].

\bibitem{Liu:2024lxy}
X.~Liu, W.~Vogelsang, F.~Yuan and H.~X. Zhu, {\it {Universality in the Near-Side Energy-Energy Correlator}},  \href{http://arXiv.org/abs/2410.16371}{{\tt arXiv:2410.16371 [hep-ph]}}.

\bibitem{Barata:2024wsu}
J.~Barata, Z.-B. Kang, X.~Mayo~L{\'o}pez and J.~Penttala, {\it {Energy-Energy Correlator for Jet Production in pp and pA Collisions}},  \href{http://dx.doi.org/10.1103/96xh-bd1w}{{\em Phys. Rev. Lett.} {\bf 134} (2025)~no.~25 251903} [\href{http://arXiv.org/abs/2411.11782}{{\tt arXiv:2411.11782 [hep-ph]}}].

\bibitem{Fu:2024pic}
Y.~Fu, B.~M\"uller and C.~Sirimanna, {\it {Modification of the Jet Energy-Energy Correlator in Cold Nuclear Matter}},  \href{http://arXiv.org/abs/2411.04866}{{\tt arXiv:2411.04866 [nucl-th]}}.

\bibitem{SLD:1994idb}
{\bf SLD} collaboration, K.~Abe {\em et.~al.}, {\it {Measurement of alpha-s (M(Z)**2) from hadronic event observables at the Z0 resonance}},  \href{http://dx.doi.org/10.1103/PhysRevD.51.962}{{\em Phys. Rev. D} {\bf 51} (1995) 962} [\href{http://arXiv.org/abs/hep-ex/9501003}{{\tt arXiv:hep-ex/9501003}}].

\bibitem{L3:1992btq}
{\bf L3} collaboration, O.~Adrian {\em et.~al.}, {\it {Determination of alpha-s from hadronic event shapes measured on the Z0 resonance}},  \href{http://dx.doi.org/10.1016/0370-2693(92)90463-E}{{\em Phys. Lett. B} {\bf 284} (1992) 471}.

\bibitem{OPAL:1991uui}
{\bf OPAL} collaboration, P.~D. Acton {\em et.~al.}, {\it {An Improved measurement of alpha-s (M (Z0)) using energy correlations with the OPAL detector at LEP}},  \href{http://dx.doi.org/10.1016/0370-2693(92)91681-X}{{\em Phys. Lett. B} {\bf 276} (1992) 547}.

\bibitem{TOPAZ:1989yod}
{\bf TOPAZ} collaboration, I.~Adachi {\em et.~al.}, {\it {Measurements of $\alpha^- s$ in $e^+ e^-$ Annihilation at $\sqrt{s}=53$.3-{GeV} and 59.5-{GeV}}},  \href{http://dx.doi.org/10.1016/0370-2693(89)90969-6}{{\em Phys. Lett. B} {\bf 227} (1989) 495}.

\bibitem{TASSO:1987mcs}
{\bf TASSO} collaboration, W.~Braunschweig {\em et.~al.}, {\it {A Study of Energy-energy Correlations Between 12-{GeV} and 46.8-{GeV} {CM} Energies}},  \href{http://dx.doi.org/10.1007/BF01573928}{{\em Z. Phys. C} {\bf 36} (1987) 349}.

\bibitem{JADE:1984taa}
{\bf JADE} collaboration, W.~Bartel {\em et.~al.}, {\it {Measurements of Energy Correlations in $e^+ e^- \to$ Hadrons}},  \href{http://dx.doi.org/10.1007/BF01547922}{{\em Z. Phys. C} {\bf 25} (1984) 231}.

\bibitem{Fernandez:1984db}
E.~Fernandez {\em et.~al.}, {\it {A Measurement of Energy-energy Correlations in $e^+ e^- \to$ Hadrons at $\sqrt{s}=29$-{GeV}}},  \href{http://dx.doi.org/10.1103/PhysRevD.31.2724}{{\em Phys. Rev. D} {\bf 31} (1985) 2724}.

\bibitem{Wood:1987uf}
D.~R. Wood {\em et.~al.}, {\it {Determination of $\alpha^- s$ From Energy-energy Correlations in $e^+ e^-$ Annihilation at 29-{GeV}}},  \href{http://dx.doi.org/10.1103/PhysRevD.37.3091}{{\em Phys. Rev. D} {\bf 37} (1988) 3091}.

\bibitem{CELLO:1982rca}
{\bf CELLO} collaboration, H.~J. Behrend {\em et.~al.}, {\it {Analysis of the Energy Weighted Angular Correlations in Hadronic $e^+ e^-$ Annihilations at 22-{GeV} and 34-{GeV}}},  \href{http://dx.doi.org/10.1007/BF01495029}{{\em Z. Phys. C} {\bf 14} (1982) 95}.

\bibitem{PLUTO:1985yzc}
{\bf PLUTO} collaboration, C.~Berger {\em et.~al.}, {\it {A Study of Energy-energy Correlations in $e^+ e^-$ Annihilations at $\sqrt{s}=34$.6-{GeV}}},  \href{http://dx.doi.org/10.1007/BF01413599}{{\em Z. Phys. C} {\bf 28} (1985) 365}.

\bibitem{OPAL:1990reb}
{\bf OPAL} collaboration, M.~Z. Akrawy {\em et.~al.}, {\it {A Measurement of energy correlations and a determination of alpha-s (M2 (Z0)) in e+ e- annihilations at s**(1/2) = 91-GeV}},  \href{http://dx.doi.org/10.1016/0370-2693(90)91098-V}{{\em Phys. Lett. B} {\bf 252} (1990) 159}.

\bibitem{ALEPH:1990vew}
{\bf ALEPH} collaboration, D.~Decamp {\em et.~al.}, {\it {Measurement of alpha-s from the structure of particle clusters produced in hadronic Z decays}},  \href{http://dx.doi.org/10.1016/0370-2693(91)91926-M}{{\em Phys. Lett. B} {\bf 257} (1991) 479}.

\bibitem{L3:1991qlf}
{\bf L3} collaboration, B.~Adeva {\em et.~al.}, {\it {Determination of alpha-s from energy-energy correlations measured on the Z0 resonance.}},  \href{http://dx.doi.org/10.1016/0370-2693(91)91925-L}{{\em Phys. Lett. B} {\bf 257} (1991) 469}.

\bibitem{SLD:1994yoe}
{\bf SLD} collaboration, K.~Abe {\em et.~al.}, {\it {Measurement of alpha-s from energy-energy correlations at the Z0 resonance}},  \href{http://dx.doi.org/10.1103/PhysRevD.50.5580}{{\em Phys. Rev. D} {\bf 50} (1994) 5580} [\href{http://arXiv.org/abs/hep-ex/9405006}{{\tt arXiv:hep-ex/9405006}}].

\bibitem{Bossi:2024qeu}
H.~Bossi, A.~Baty, Y.~Chen, Y.-C.~J. Chen, G.-M. Innocenti, M.~Maggi, C.~McGinn and Y.-J. Lee, {\it {Measurement of the energy-energy correlator in the back-to-back limit using the archived ALEPH e+e- data at 91.2 GeV}},  \href{http://dx.doi.org/10.22323/1.478.0228}{{\em PoS} {\bf LHCP2024} (2025) 228} [\href{http://arXiv.org/abs/2501.01968}{{\tt arXiv:2501.01968 [hep-ex]}}].

\bibitem{Bossi:2025xsi}
H.~Bossi, Y.-C. Chen, Y.~Chen, J.~Zhang, G.~M. Innocenti, A.~Badea, A.~Baty, M.~Maggi, C.~McGinn and Y.-J. Lee, {\it {Analysis note: measurement of energy-energy correlator in $e^{+}e^{-}$ collisions at $91$ GeV with archived ALEPH data}},  \href{http://arXiv.org/abs/2505.11828}{{\tt arXiv:2505.11828 [hep-ex]}}.

\bibitem{CMS:2024mlf}
{\bf CMS} collaboration, A.~Hayrapetyan {\em et.~al.}, {\it {Measurement of Energy Correlators inside Jets and Determination of the Strong Coupling \ensuremath{\alpha}S(mZ)}},  \href{http://dx.doi.org/10.1103/PhysRevLett.133.071903}{{\em Phys. Rev. Lett.} {\bf 133} (2024)~no.~7 071903} [\href{http://arXiv.org/abs/2402.13864}{{\tt arXiv:2402.13864 [hep-ex]}}].

\bibitem{ALICE:2024dfl}
{\bf ALICE} collaboration, S.~Acharya {\em et.~al.}, {\it {Exposing the parton-hadron transition within jets with energy-energy correlators in pp collisions at $\sqrt{\textit s}=5.02$ TeV}},  \href{http://arXiv.org/abs/2409.12687}{{\tt arXiv:2409.12687 [hep-ex]}}.

\bibitem{Komiske:2022enw}
P.~T. Komiske, I.~Moult, J.~Thaler and H.~X. Zhu, {\it {Analyzing N-Point Energy Correlators inside Jets with CMS Open Data}},  \href{http://dx.doi.org/10.1103/PhysRevLett.130.051901}{{\em Phys. Rev. Lett.} {\bf 130} (2023)~no.~5 051901} [\href{http://arXiv.org/abs/2201.07800}{{\tt arXiv:2201.07800 [hep-ph]}}].

\bibitem{Lee:2025okn}
K.~Lee and I.~Stewart, {\it {Dihadron Fragmentation and the Confinement Transition in Energy Correlators}},  \href{http://arXiv.org/abs/2507.11495}{{\tt arXiv:2507.11495 [hep-ph]}}.

\bibitem{Guo:2025zwb}
Y.~Guo, F.~Yuan and W.~Zhao, {\it {Factorization and Resummation for the Nearside Energy-Energy Correlators}},  \href{http://arXiv.org/abs/2507.15820}{{\tt arXiv:2507.15820 [hep-ph]}}.

\bibitem{Chang:2025kgq}
C.-H. Chang, H.~Chen, X.~Liu, D.~Simmons-Duffin, F.~Yuan and H.~X. Zhu, {\it {Quantum Scaling in Energy Correlators Beyond the Confinement Transition}},  \href{http://arXiv.org/abs/2507.15923}{{\tt arXiv:2507.15923 [hep-ph]}}.

\bibitem{Kang:2025EEC}
Z.-B. Kang, A.~Metz, D.~Pitonyak and C.~Zhang, {\it {Dihadron fragmentation framework for near-side energy-energy correlators}},  \href{http://arXiv.org/abs/2507.xxxxx}{{\tt arXiv:2507.xxxxx [hep-ph]}}.
\newblock to appear.

\bibitem{Korchemsky:1999kt}
G.~P. Korchemsky and G.~F. Sterman, {\it {Power corrections to event shapes and factorization}},  \href{http://dx.doi.org/10.1016/S0550-3213(99)00308-9}{{\em Nucl. Phys. B} {\bf 555} (1999) 335} [\href{http://arXiv.org/abs/hep-ph/9902341}{{\tt arXiv:hep-ph/9902341}}].

\bibitem{Lee:2024esz}
K.~Lee, A.~Pathak, I.~W. Stewart and Z.~Sun, {\it {Nonperturbative Effects in Energy Correlators: From Characterizing Confinement Transition to Improving {\ensuremath{\alpha}}s Extraction}},  \href{http://dx.doi.org/10.1103/PhysRevLett.133.231902}{{\em Phys. Rev. Lett.} {\bf 133} (2024)~no.~23 231902} [\href{http://arXiv.org/abs/2405.19396}{{\tt arXiv:2405.19396 [hep-ph]}}].

\bibitem{Dixon:2018qgp}
L.~J. Dixon, M.-X. Luo, V.~Shtabovenko, T.-Z. Yang and H.~X. Zhu, {\it {Analytical Computation of Energy-Energy Correlation at Next-to-Leading Order in QCD}},  \href{http://dx.doi.org/10.1103/PhysRevLett.120.102001}{{\em Phys. Rev. Lett.} {\bf 120} (2018)~no.~10 102001} [\href{http://arXiv.org/abs/1801.03219}{{\tt arXiv:1801.03219 [hep-ph]}}].

\bibitem{DelDuca:2016csb}
V.~Del~Duca, C.~Duhr, A.~Kardos, G.~Somogyi and Z.~Tr\'ocs\'anyi, {\it {Three-Jet Production in Electron-Positron Collisions at Next-to-Next-to-Leading Order Accuracy}},  \href{http://dx.doi.org/10.1103/PhysRevLett.117.152004}{{\em Phys. Rev. Lett.} {\bf 117} (2016)~no.~15 152004} [\href{http://arXiv.org/abs/1603.08927}{{\tt arXiv:1603.08927 [hep-ph]}}].

\bibitem{Tulipant:2017ybb}
Z.~Tulip\'ant, A.~Kardos and G.~Somogyi, {\it {Energy\textendash{}energy correlation in electron\textendash{}positron annihilation at NNLL + NNLO accuracy}},  \href{http://dx.doi.org/10.1140/epjc/s10052-017-5320-9}{{\em Eur. Phys. J. C} {\bf 77} (2017)~no.~11 749} [\href{http://arXiv.org/abs/1708.04093}{{\tt arXiv:1708.04093 [hep-ph]}}].

\bibitem{Kalinowski:1980wea}
J.~Kalinowski, K.~Konishi, P.~N. Scharbach and T.~R. Taylor, {\it {RESOLVING QCD JETS BEYOND LEADING ORDER: QUARK DECAY PROBABILITIES}},  \href{http://dx.doi.org/10.1016/0550-3213(81)90352-7}{{\em Nucl. Phys. B} {\bf 181} (1981) 253}.

\bibitem{RICHARDS1982193}
D.~Richards, W.~Stirling and S.~Ellis, {\it Second order corrections to the energy-energy correlation function in quantum chromodynamics},  \href{http://dx.doi.org/https://doi.org/10.1016/0370-2693(82)90275-1}{{\em Physics Letters B} {\bf 119} (1982)~no.~1 193}.
\newblock \url{https://www.sciencedirect.com/science/article/pii/0370269382902751}.

\bibitem{Collins:2011zzd}
J.~Collins, {\em {Foundations of Perturbative QCD}}, vol.~32 of {\em Cambridge Monographs on Particle Physics, Nuclear Physics and Cosmology}.
\newblock Cambridge University Press, 7, 2023.

\bibitem{Rijken:1996ns}
P.~J. Rijken and W.~L. van Neerven, {\it {Higher order QCD corrections to the transverse and longitudinal fragmentation functions in electron - positron annihilation}},  \href{http://dx.doi.org/10.1016/S0550-3213(96)00669-4}{{\em Nucl. Phys. B} {\bf 487} (1997) 233} [\href{http://arXiv.org/abs/hep-ph/9609377}{{\tt arXiv:hep-ph/9609377}}].

\bibitem{Mitov:2006wy}
A.~Mitov and S.-O. Moch, {\it {QCD Corrections to Semi-Inclusive Hadron Production in Electron-Positron Annihilation at Two Loops}},  \href{http://dx.doi.org/10.1016/j.nuclphysb.2006.05.018}{{\em Nucl. Phys. B} {\bf 751} (2006) 18} [\href{http://arXiv.org/abs/hep-ph/0604160}{{\tt arXiv:hep-ph/0604160}}].

\bibitem{Mitov:2006ic}
A.~Mitov, S.~Moch and A.~Vogt, {\it {Next-to-Next-to-Leading Order Evolution of Non-Singlet Fragmentation Functions}},  \href{http://dx.doi.org/10.1016/j.physletb.2006.05.005}{{\em Phys. Lett. B} {\bf 638} (2006) 61} [\href{http://arXiv.org/abs/hep-ph/0604053}{{\tt arXiv:hep-ph/0604053}}].

\bibitem{Moch:2007tx}
S.~Moch and A.~Vogt, {\it {On third-order timelike splitting functions and top-mediated Higgs decay into hadrons}},  \href{http://dx.doi.org/10.1016/j.physletb.2007.10.069}{{\em Phys. Lett. B} {\bf 659} (2008) 290} [\href{http://arXiv.org/abs/0709.3899}{{\tt arXiv:0709.3899 [hep-ph]}}].

\bibitem{Almasy:2011eq}
A.~A. Almasy, S.~Moch and A.~Vogt, {\it {On the Next-to-Next-to-Leading Order Evolution of Flavour-Singlet Fragmentation Functions}},  \href{http://dx.doi.org/10.1016/j.nuclphysb.2011.08.028}{{\em Nucl. Phys. B} {\bf 854} (2012) 133} [\href{http://arXiv.org/abs/1107.2263}{{\tt arXiv:1107.2263 [hep-ph]}}].

\bibitem{Collins:1984kg}
J.~C. Collins, D.~E. Soper and G.~F. Sterman, {\it {Transverse Momentum Distribution in Drell-Yan Pair and W and Z Boson Production}},  \href{http://dx.doi.org/10.1016/0550-3213(85)90479-1}{{\em Nucl. Phys. B} {\bf 250} (1985) 199}.

\bibitem{Collins:2014jpa}
J.~Collins and T.~Rogers, {\it {Understanding the large-distance behavior of transverse-momentum-dependent parton densities and the Collins-Soper evolution kernel}},  \href{http://dx.doi.org/10.1103/PhysRevD.91.074020}{{\em Phys. Rev. D} {\bf 91} (2015)~no.~7 074020} [\href{http://arXiv.org/abs/1412.3820}{{\tt arXiv:1412.3820 [hep-ph]}}].

\bibitem{Apolinario:2025vtx}
L.~Apolin{\'a}rio, R.~Kunnawalkam~Elayavalli, N.~O. Madureira, J.-X. Sheng, X.-N. Wang and Z.~Yang, {\it {Flavor dependence of Energy-energy correlators}},  \href{http://arXiv.org/abs/2502.11406}{{\tt arXiv:2502.11406 [hep-ph]}}.

\bibitem{Kardos:2018kqj}
A.~Kardos, S.~Kluth, G.~Somogyi, Z.~Tulip\'ant and A.~Verbytskyi, {\it {Precise determination of $\alpha _{S}(M_Z)$ from a global fit of energy\textendash{}energy correlation to NNLO+NNLL predictions}},  \href{http://dx.doi.org/10.1140/epjc/s10052-018-5963-1}{{\em Eur. Phys. J. C} {\bf 78} (2018)~no.~6 498} [\href{http://arXiv.org/abs/1804.09146}{{\tt arXiv:1804.09146 [hep-ph]}}].

\bibitem{replica-spin-class}
M.~Mezard, G.~Parisi and M.~Virasoro, {\em Spin Glass Theory and Beyond}.
\newblock WORLD SCIENTIFIC, 1986.

\bibitem{Ball:2008by}
{\bf NNPDF} collaboration, R.~D. Ball, L.~Del~Debbio, S.~Forte, A.~Guffanti, J.~I. Latorre, A.~Piccione, J.~Rojo and M.~Ubiali, {\it {A Determination of parton distributions with faithful uncertainty estimation}},  \href{http://dx.doi.org/10.1016/j.nuclphysb.2008.09.037}{{\em Nucl. Phys. B} {\bf 809} (2009) 1} [\href{http://arXiv.org/abs/0808.1231}{{\tt arXiv:0808.1231 [hep-ph]}}].
\newblock [Erratum: Nucl.Phys.B 816, 293 (2009)].

\bibitem{Bell:2023dqs}
G.~Bell, C.~Lee, Y.~Makris, J.~Talbert and B.~Yan, {\it {Effects of renormalon scheme and perturbative scale choices on determinations of the strong coupling from e+e- event shapes}},  \href{http://dx.doi.org/10.1103/PhysRevD.109.094008}{{\em Phys. Rev. D} {\bf 109} (2024)~no.~9 094008} [\href{http://arXiv.org/abs/2311.03990}{{\tt arXiv:2311.03990 [hep-ph]}}].

\bibitem{Jaarsma:2023ell}
M.~Jaarsma, Y.~Li, I.~Moult, W.~J. Waalewijn and H.~X. Zhu, {\it {Energy correlators on tracks: resummation and non-perturbative effects}},  \href{http://dx.doi.org/10.1007/JHEP12(2023)087}{{\em JHEP} {\bf 12} (2023) 087} [\href{http://arXiv.org/abs/2307.15739}{{\tt arXiv:2307.15739 [hep-ph]}}].

\bibitem{Lee:2023npz}
K.~Lee and I.~Moult, {\it {Energy Correlators Taking Charge}},  \href{http://arXiv.org/abs/2308.00746}{{\tt arXiv:2308.00746 [hep-ph]}}.

\bibitem{Lee:2023tkr}
K.~Lee and I.~Moult, {\it {Joint Track Functions: Expanding the Space of Calculable Correlations at Colliders}},  \href{http://arXiv.org/abs/2308.01332}{{\tt arXiv:2308.01332 [hep-ph]}}.

\bibitem{STAR:2025jut}
{\bf STAR} collaboration, {\it {Measurement of Two-Point Energy Correlators Within Jets in $p$+$p$ Collisions at $\sqrt{s}$ = 200 GeV}},  \href{http://arXiv.org/abs/2502.15925}{{\tt arXiv:2502.15925 [hep-ex]}}.

\end{thebibliography}\endgroup

\clearpage
\onecolumngrid
\section*{Supplementary material}

\subsection{Factorization of EEC in the Collinear Limit}

In this section, we review the collinear factorization of the EEC in momentum space for $e^+e^-$ annihilation, following Ref.~\cite{Dixon:2019uzg}. Our goal is to lay the foundation for the position-space factorization introduced in the next section, which will be essential for incorporating non-perturbative effects. 

The \emph{cumulant EEC} is defined as
\begin{align}
\Sigma(z,Q)
&=
\int_0^z \dd{z'} \, \frac{\dd{\Sigma}}{\dd{z'}}(z',Q)\, ,
\end{align}
where the EEC is related to Eq.~(2) in the main text via
\begin{align}
 \frac{\dd{\Sigma}}{\dd{z}} 
 = \frac{\dd \chi }{\dd z} \frac{\dd{\Sigma}}{\dd{\chi}} 
 = \frac{1}{\sqrt{z(1-z)}} \sum_{i,j} \int \dd{\sigma_{ij} }
 \frac{E_i \, E_j}{Q^2} \delta(\chi -\theta_{ij})
 = \sum_{i,j} \int  \dd{\sigma_{ij} } \frac{E_i \, E_j}{Q^2} \, \delta \!\left(z- \frac{1}{2}(1-\cos \theta_{ij})\right)
 \,.
\end{align}
The cumulant EEC $\Sigma(z,Q)$ factorizes into a convolution of the two-component EEC jet function $ \mathbf{J}=(J_q,J_g)$ \cite{Dixon:2019uzg} and the hard function $\mathbf{ H}=(H_q,H_g)$ \cite{Rijken:1996ns, Mitov:2006wy, Mitov:2006ic,Moch:2007tx,Almasy:2011eq}:
\begin{align}
\label{eq:EEC_mom}
\Sigma(z,Q)
&=\int_0^1 \dd{x} \,x^2\,
\mathbf{J}\qty(x\,q_T,\mu)\vdot
\mathbf{H}\qty(x,Q,\mu)\,.
\end{align}
Here, we use slightly simplified notation compared to Ref.~\cite{Dixon:2019uzg}, writing $\Sigma(z,\log(Q^2/\mu^2),\mu)$ as $\Sigma(z,Q)$ for brevity.
In the jet function $\mathbf{J}$, the transverse momentum scale $x q_T = x \sqrt{z}\, Q = x Q \sin{\frac{\chi}{2} }$ has the physical interpretation as the relative transverse momentum between the pair of final state hadrons. As the jet function factorizes from the hard collision, it is process-independent. The hard function obeys the usual timelike DGLAP evolution~\cite{Dixon:2019uzg}:
\begin{align}
\label{eq:DGLAP}
\frac{\dd{\mathbf{H}(x,Q,\mu)}}{\dd{\ln\mu^2}}
&= -\int_x^1 \frac{\dd{y}}{y}\,\mathbf{P}(y, \mu)\,\vdot \mathbf{H}\qty(\frac{x}{y},Q,\mu)\,.
\end{align}
The fixed-order (\emph{non-resummed}) hard functions, i.e., the initial conditions for the DGLAP evolution, can be computed perturbatively and contain logarithms that are minimized at the canonical hard scale $\mu_H = Q$. The timelike splitting kernel matrix is given by
\begin{equation}
\mathbf{P}
= \begin{pmatrix}
P_{qq} & P_{qg}\\
P_{gq} & P_{gg}
\end{pmatrix}\,,
\end{equation}
where $P_{ij}$ represents the probability for the parton branching $j \rightarrow i + X$. Up to NNLO, the timelike splitting kernels can be found in Refs.~\cite{Rijken:1996ns, Mitov:2006wy, Mitov:2006ic,Moch:2007tx,Almasy:2011eq}. Requiring $\Sigma$ to be independent of $\mu$ implies an analogous equation for the jet function~\cite{Dixon:2019uzg}:
\begin{align}
\label{eq:J_mom_RGE}
\frac{\dd{\mathbf{J}(p,\mu)}}{\dd{\ln\mu^2}}
&= \int_0^1 \dd{y}\,y^2\, \mathbf{J}(y p,\mu) \vdot \mathbf{P}(y, \mu) \,,
\end{align}
with the same timelike splitting kernel matrix $\mathbf{P}$.

\subsection{Collinear Factorization of EEC in Position Space}
We now describe the factorization of the EEC in position space. Define the \emph{differential} jet function in momentum space:
\begin{align}
\label{eq:diff_J}
\mathbf{j}(p,\mu)
&=\frac{1}{2\pi p}\frac{\dd}{\dd{p}}\,\mathbf{J}(p,\mu)\,,
\end{align}
whose Fourier transform is
\begin{align} 
\label{eq:pos_J}
\widetilde{\mathbf{j}}(b,\mu)
&=\int \dd[2]{\vec p} \;e^{\,i\vec p\cdot\vec b}\;
\mathbf{j}(p,\mu)\, .
\end{align}
The EEC then factorizes in position space as
\begin{align}
\label{eq:col_EEC_pos_space_factorization}
\frac{\dd{\Sigma(z,Q)}}{\dd{z}}
=\int_0^1 \dd{x}\,x^2 
\left[\pi Q^2 x^2 \mathbf{j} \qty(x \sqrt{z} Q,\mu)\right]
\vdot \mathbf{H}(x,Q,\mu)
=\frac{Q^2}{2}\int_0^\infty \dd{b} b\;
  J_0(b \sqrt{z}Q)\;\widetilde\Sigma(b,Q)\,,
\end{align}
where
\begin{align}
\label{eq:pos_space_EEC}
\widetilde\Sigma(b,Q)
&=\int_0^1 \dd{x}\,x^2\,
  \widetilde{\mathbf{j}} \qty(\frac{b}{x},\mu)\vdot
 \mathbf{H}(x,Q,\mu)\,,
\end{align}
and $J_0$ is the zeroth-order Bessel function of the first kind.

From Eq.~\eqref{eq:col_EEC_pos_space_factorization} we can also derive an evolution equation for the jet function in position space by demanding that
\begin{equation}
\label{eq:EEC_mu_dep}
    \frac{\dd}{\dd{\ln \mu^2}} 
\frac{\dd{\Sigma(z,Q)}}{\dd{z}} = 0.
\end{equation}
As the hard function is independent of the collinear momentum scale $q_T$, it is not modified by the transverse Fourier transformation to position space and still satisfies the same timelike DGLAP evolution~\eqref{eq:DGLAP}. Together with Eq.~\eqref{eq:EEC_mu_dep}, this implies that the position-space jet function must satisfy
\begin{align}
\label{eq:J_pos_RGE}
\frac{\dd}{\dd{\ln\mu^{2}}}\,
\widetilde{\mathbf{j}}(b,\mu)
&=\int_{0}^{1}\dd{y}\,y^{2}\,
   \widetilde{\mathbf{j}}\qty(\frac{b}{y},\mu)\vdot
   \mathbf{P}(y, \mu)\, .
\end{align}
Notably, this has the same form as the momentum-space jet evolution in Eq.~\eqref{eq:J_mom_RGE} with the replacement $yp \to b/y$ inside the jet function, and the splitting kernel matrix $\mathbf{P}$ is not modified.

%
\subsection{Initial Condition for the EEC Jet Function in Position Space}

For small values of $b$, corresponding to large momenta, we can calculate the position-space jet function perturbatively. This will serve as the perturbative initial condition to the evolution equation~\eqref{eq:J_pos_RGE}. The calculation follows that of the momentum-space jet function~\cite{Dixon:2019uzg}, except that before expanding in $\varepsilon$ in dimensional regularization, one first has to compute the position-space jet function $\widetilde{\mathbf{j}}$ using Eqs.~\eqref{eq:diff_J} and \eqref{eq:pos_J}. This results in the following general form for the position-space jet function:
\begin{align}
\label{eq:J_pos_expansion}
\widetilde j_{i}(b,\mu)
= 
\sum_{n=0} a^n_s(\mu) \sum_{m=0}^n  \widetilde j_{i, n}^{[m]} L_b^m\,,
\end{align}
where $i =q$ or $g$, and we have denoted $a_s(\mu) = \alpha_s(\mu)/(4\pi)$, $L_b = \ln{\frac{b^2 \mu^2}{b_0^2}}$, and $b_0 =2e^{-\gamma_E}$. The constants $\widetilde j_{i,n}^{[m]}$ can be computed perturbatively. This can be compared to the similar expansion in momentum space:
\begin{equation}
\label{eq:J_mom_expansion}
    J_{i}(p,\mu)
= 
\sum_{n=0} a^n_s(\mu) \sum_{m=0}^n  J_{i, n}^{[m]} L_p^m\,,
\end{equation}
where $L_p=\ln{\frac{\mu^2}{p^2}}$, and the constants $J_{i, n}^{[m]}$ have been computed at NNLO in Ref.~\cite{Dixon:2019uzg}.

To determine the constants $\widetilde j_{i,n}^{[m]}$, we note that both momentum-space and position-space expressions for the EEC, Eqs.~\eqref{eq:EEC_mom} and \eqref{eq:col_EEC_pos_space_factorization}, have to produce the same fixed-order result. By comparing these two expressions for $\frac{\dd{\Sigma}}{\dd{z}}$ at the NNLO accuracy, we find that
\begin{align}
\widetilde j^{[m]}_{i,n} = J^{[m]}_{i,n},
\end{align}
i.e., the coefficients in the perturbative expansions~\eqref{eq:J_pos_expansion} and \eqref{eq:J_mom_expansion} match up to $n \leq 2$.

\subsection{Distribution of parameters}
    \begin{figure*}[htb]
    \centering
    \includegraphics[width =0.9\textwidth]{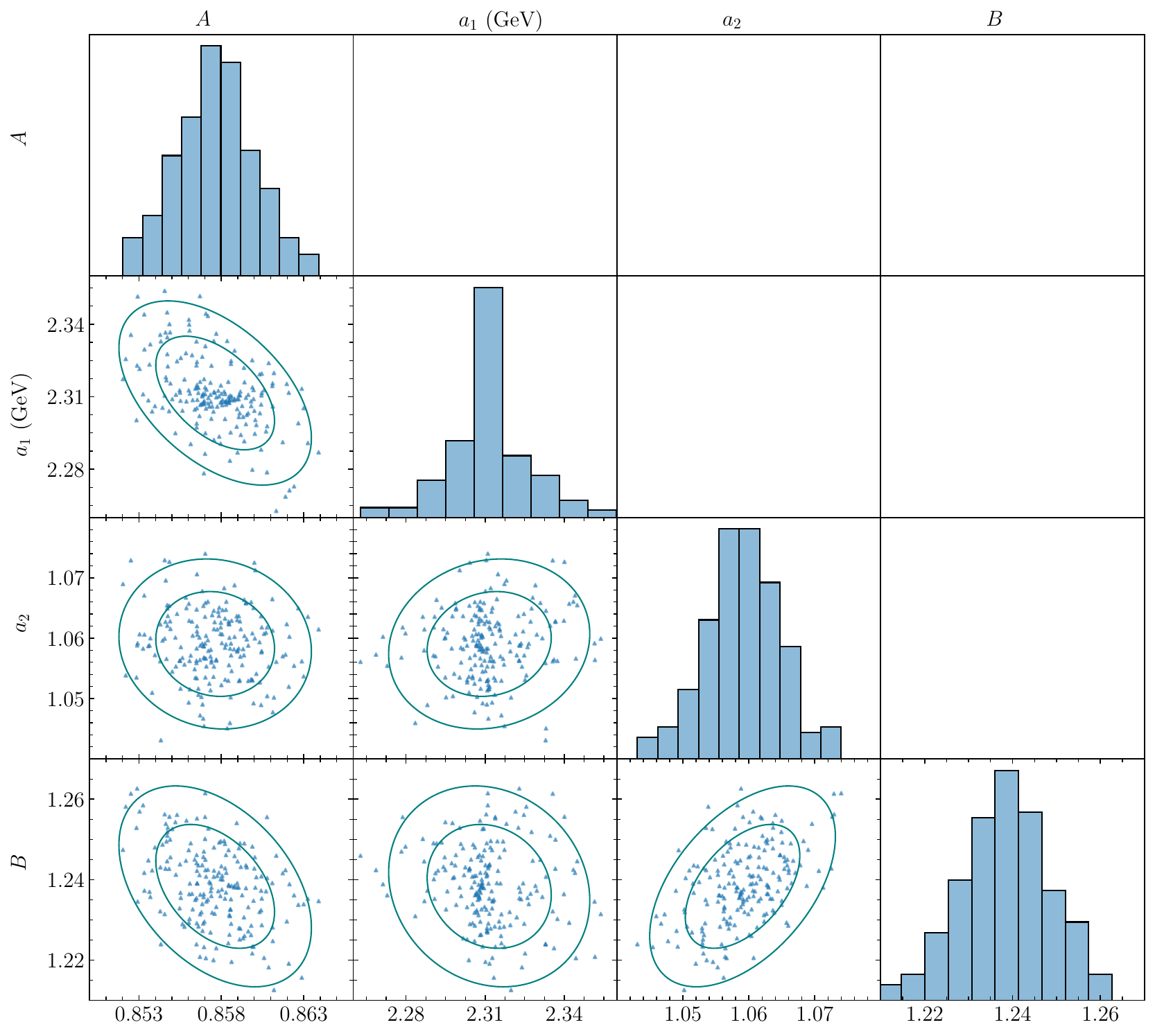}
        \caption{The distributions of the fit parameters for the 200 fit replicas. The inner and outer error ellipses correspond to the $68 \%$ and $95 \%$ confidence levels, respectively.}
    \label{distribution}
    \end{figure*}
In this work, we extracted two $K$-factors---$A$ for the resummed contribution, $\frac{\dd\Sigma^{\mathrm{res}}}{\dd z}$, and $B$ for the non-singular contribution, $\frac{\dd\Sigma^{\mathrm{nons}}}{\dd z}$, to the EEC---along with a non-perturbative (NP) modification to the jet function, given by
\begin{equation}
  j_{\mathrm{np}}(b) = \exp\!\left[-(a_{1} b)^{a_{2}}\right].
\end{equation}
As discussed in the main text, the uncertainties on the fitted parameters were estimated using the replica method: Gaussian noise was added to the data to generate multiple replicas, and all parameters were refitted independently for each replica.

Figure~\ref{distribution} shows the distributions of the resulting parameters. The diagonal panels give the one-dimensional density distributions; off-diagonal panels show the fitted values and the confidence ellipses at 68\% and 95\% against two given parameters. The absolute correlation between any pair of parameters is below $0.5$.

\end{document}